\documentclass[aps,twocolumn,amsmath,amssymb,floatfix]{revtex4}
\usepackage{dcolumn}
\usepackage{bm}
\usepackage{amsfonts}
\usepackage{graphicx}  % Include figure files
\usepackage{dcolumn}  % Align table columns on decimal point
\usepackage{bm}  % bold math
\usepackage{amsmath}  % needed for subequations
\usepackage{amssymb}
\usepackage{multirow}
\usepackage{booktabs}
\usepackage{xcolor}
\usepackage{float}
\usepackage{titlesec}

\renewcommand{\thesection}{\arabic{section}} 
\renewcommand{\thesubsection}{\thesection.\arabic{subsection}}

\titleformat{\section}[block]{\large\bf}{\thesection.}{1em}{}
\titleformat{\subsection}[block]{\bf}{\thesubsection.}{1em}{}

\begin{document}
\title{Broken sublattice symmetry states in Bernal stacked multilayer graphene}
\author{Chiho Yoon$^{1}$}
\author{Yunsu Jang$^{1}$}
\author{Jeil Jung$^{2}$}
\email{jeil.jung@gmail.com}
\author{Hongki Min$^{1}$}
\email{hmin@snu.ac.kr}
\affiliation{$^1$ Department of Physics and Astronomy, Seoul National University, Seoul 08826, Korea}
\affiliation{$^2$ Department of Physics, University of Seoul, Seoul 02504, Korea}

\date{\today}

\begin{abstract}
We analyze the ordered phases of Bernal stacked multilayer graphene in the presence of interaction induced band gaps due to sublattice symmetry breaking potentials, whose solutions can be analyzed in terms of light-mass and heavy-mass pseudospin doublets which have the same Chern numbers but opposite charge polarization directions. The application of a perpendicular external electric field reveals an effective Hund's rule for the ordering of the sublattice pseudospin doublets in a tetralayer, while a similar but more complex phase diagram develops with increasing layer number.

\end{abstract}

\maketitle
%\normalsize

%%%%%%%%%%%%%%%%%%%%%%%%%%%%%%%%%%%%%%%%%%%%%%%%%%%%%%%%%%%%%%%%%%%%%%%%%%%%%%%%%%%%%%%%%%%%%%%%%%%%
\section {Introduction}
Ultrathin multilayer graphene has been extensively studied in the literature over the last decade as a promising platform for electronic devices %\cite{device_application} 
\cite{CastroNeto2009,DasSarma2011,Basov2014,HassanRaza2012}
and energy storage applications \cite{ruoff} that take advantage of the superlative properties of graphene. 
From a more fundamental physics point of view, few-layer graphenes are interesting because their band structure
embodies the chiral nature of the Dirac cones near the charge neutrality point which can manifest in transport and optical experiments.
Clear signatures of electron-electron interactions observed 
through scanning probes \cite{yacoby,Feldman2009} and transport experiments \cite{geimnematic,lau,freitag,Veligura2012} 
have signaled interesting many-body effects.
Remarkably, the predictions of interaction driven band gaps in Bernal stacked bilayer \cite{Min2008a} and rhombohedral trilayer graphene \cite{jung2013} have been speculated to be accompanied by spin/valley resolved spontaneous Hall phases \cite{rahoul,Zhang2011,jung2011,Zhang2015} for a variety of possible ground-state configurations among quasi-degenerate states.
Other possible ordered phases suggested near the charge neutrality point in bilayer graphene include nematic phases with broken rotational symmetry \cite{intblg1,intblg2,intblg3a,intblg3b,intblg4,intblg5,geimnematic}, and Fermi surface instabilities in both $\ell=0, 1$ channels in the presence of a finite carrier doping and electric fields \cite{jung2015}.
Recent experiments in ultraclean Bernal stacked multilayer graphenes signal the formation of electron-electron interaction driven ordered phases \cite{Grushina2015, Nam2016}.

In this paper, we analyze the nature of the electron interaction driven 
ordered ground-state phases in Bernal stacked tetralayer graphene subject to perpendicular external electric fields and the associated Hall conductivities that can be measured in transport experiments.
We show that the electronic structure consisting of light-mass and heavy-mass band doublets follows an effective Hund's rule of the sublattice pseudospins when a perpendicular external electric field is applied, allowing  to introduce qualitative changes in the associated Hall conductivities. 
Interestingly, in a certain range of electric fields, a ground state with a non-vanishing charge Hall conductivity appears that should be measurable by conventional Hall experiments.  
Analysis of the ordered phases in Bernal stacked multilayer graphene beyond tetralayer acquires a more complex character due to the appearance of additional pseudospin doublets and mixing between them. 
%Within a minimal multiband model, we show that the interaction driven band gap and broken sublattice symmetry can appear Bernal stacked multilayer graphene. 
Within a minimal multiband model, we show that the interaction driven band gap and broken sublattice symmetry can appear in even-layer graphenes, whereas the gaps for odd-layer graphenes are  suppressed, exhibiting an even-odd effect for the energy gap size.
%%%%%%%%%%%%%%%%%%%%%%%%%%%%%%%%%%%%%%%%%%%%%%%%%%%%%%%%%%%%%%%%%%%%%%%%%%%%%%%%%%%%%%%%%%%%%%%%%%%%

%%%%%%%%%%%%%%%%%%%%%%%%%%%%%%%%%%%%%%%%%%%%%%%%%%
\begin{figure*}[t]
\frame{\includegraphics[width=\linewidth]{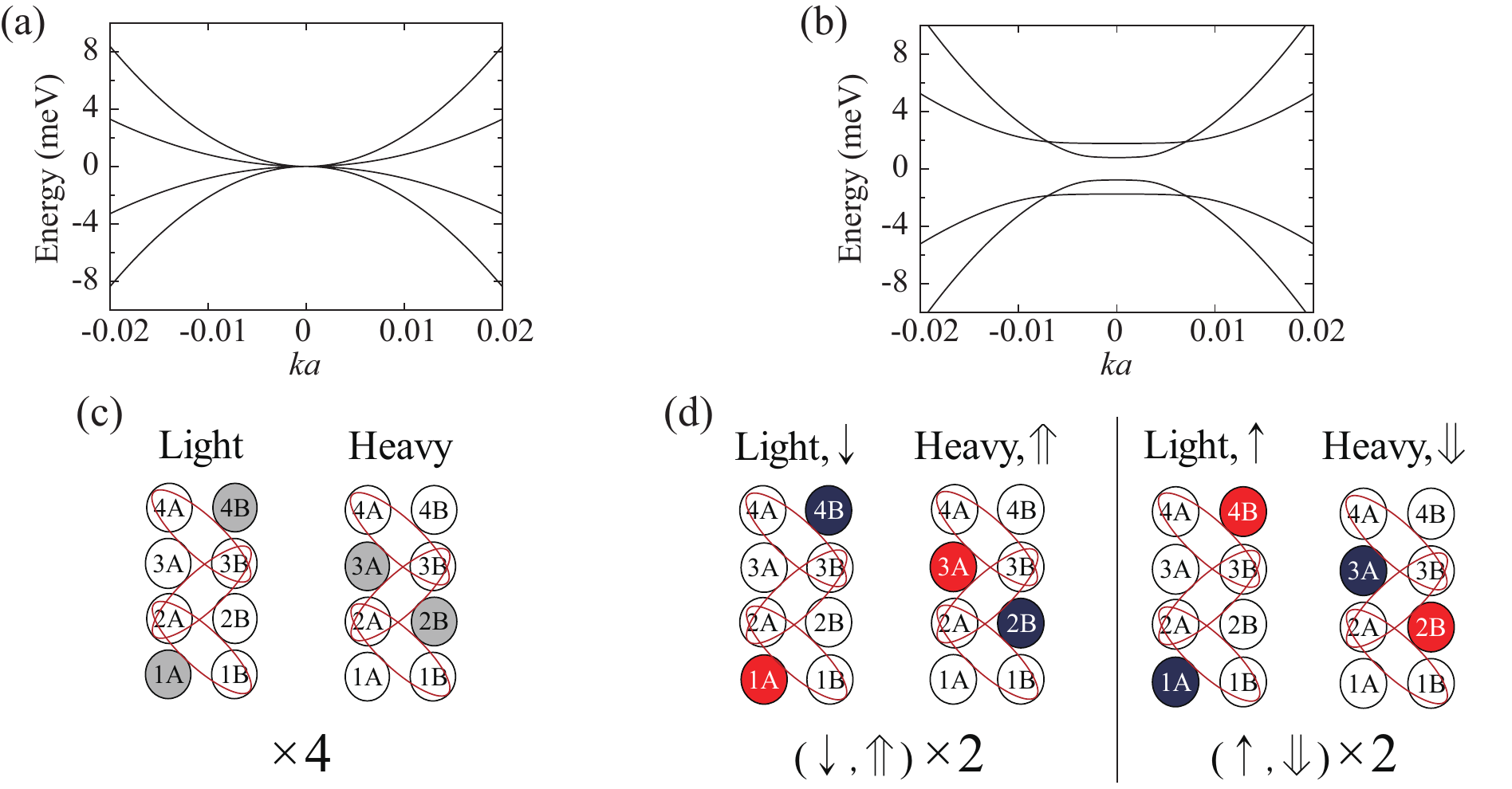}}
\caption{
Electronic structure and zero-energy wavefunction configurations near the $K$ or $K'$ valley for ABAB tetralayer graphene obtained respectively from (a), (c) the non-interacting continuum model and (b), (d) a self-consistent Hartree-Fock calculation. 
Two pseudospin doublets are labeled by ``Light" or ``Heavy'' depending on their effective mass of the energy band. In the case of non-interacting model, all four spin/valley flavors have the same wavefunction configuration with localized wavefunctions on the gray sublattices, as shown in (c). When electron-electron interactions are turned on, the sublattice symmetry is broken for both doublets transferring charges either from A to B sublattices or vice versa, as indicated in red (positive charge) and blue (negative charge) color in (d). 
%the ground-state wavefunction configurations are divided by two kinds and each kind contains two flavors as shown in (d). The blue (red) color in (d) indicates larger (smaller) wavefunction amplitude compared to the non-interacting case. 
The $(\downarrow,\Uparrow)$ and $(\uparrow,\Downarrow)$ labels represent two possible configurations of opposite charge polarization towards the top and bottom layers corresponding to the light and heavy mass bands.
} 
\label{fig:ground_state}
\end{figure*}
%%%%%%%%%%%%%%%%%%%%%%%%%%%%%%%%%%%%%%%%%%%%%%%%%%

\section {Method}
We use a $\pi$-band minimal continuum model for multilayer graphene in which only nearest-neighbor intralayer hopping $t_0$ and interlayer hopping $t_1$ for the full $\pi$-bands are retained. The non-interacting Hamiltonian  is
\begin{eqnarray}
\label{eq:TB}
\hat{H}_{0} = \sum_{\bm{k}, \sigma, \sigma'} \hat{c}_{\bm{k},\sigma}^{\dagger} \varepsilon_{\sigma \sigma'}^{(0)} (\bm{k}) \hat{c}_{\bm{k}, \sigma'}, 
\end{eqnarray}
where $\bm {k}$ is the wavevector measured from a valley $K$ or $K'$, $\sigma$ is a collective index representing spin (u/d), valley, sublattice (A/B), and layer ($n=1,2,\cdots$) degrees of freedom, $\hat{c}_{\bm{k},\sigma}^{\dagger}$ ($\hat{c}_{\bm{k},\sigma}$) is the electron creation (annihilation) operator for $\bm k$ and $\sigma$, and $\varepsilon_{\sigma\sigma'}^{(0)}(\bm{k})$ is the non-interacting Hamiltonian matrix element for %the Bernal stacking. 
Bernal stacked multilayers. % Jeil
For the tight-binding parameters, we use the { LDA parameters of graphite} $t_{0}=2.598$ eV and $t_{1}=0.377$ eV \cite{Charlier1991, Jung2014}. 

We include the effect of electron-electron interactions within a mean-field Hartree-Fock approximation, 
\begin{equation}
\hat{H}_{\rm MF}= \hat{H}_{0} + \sum_{{\bm k},\sigma,\sigma'} \hat{c}_{{\bm k},\sigma}^{\dagger} \varepsilon_{\sigma\sigma'}^{({\rm HF})}({\bm k})  \hat{c}_{{\bm k},\sigma'}.
\end{equation}
The matrix element of the Hartree-Fock term is given by
\begin{eqnarray} \label{eq:hf_equation}
\nonumber\varepsilon_{\sigma\sigma'}^{({\rm HF})}({\bm k})=\delta_{\sigma\sigma'} \sum_{{\bm k'},\sigma'} V_{nn'} (0) \left\langle \hat{c}_{{\bm k'}, \sigma'}^{\dagger} \hat{c}_{{\bm k'}, \sigma'} \right\rangle \\ 
- \delta_{s s'}\sum_{{\bm k'}} V_{nn'} (|\bm{k}-\bm{k'}|) \left\langle \hat{c}_{{\bm k'}, \sigma'}^{\dagger} \hat{c}_{{\bm k'}, \sigma} \right\rangle,
\end{eqnarray}
where $n$ and $s$ denote the layer and spin, respectively. 
$V_{nn'}(q) = \frac{2\pi e^{2} } { \epsilon_{r}q} e^{-|n-n'|qd}$ is the Coulomb interaction matrix where $d=3.35$ $\rm \AA$ is the interlayer separation and $\epsilon_r$ is the background dielectric constant.
The first and second terms in the right-hand side of Eq.~(\ref{eq:hf_equation}) represent the classical Hartree and exchange Fock contributions, respectively. 
Note that the Hartree terms reduce to potential differences between the layers when we take the proper limit at $q=0$. 
%Here we take a rather large value of $\epsilon_{r}=8$ or equivalently the interaction strength $\alpha\equiv {e^2\over \epsilon_r \hbar v}=...$ 
Here we take a rather small value of the interaction strength $\alpha\equiv {e^2\over \epsilon_r \hbar v}$, where $v=\frac{\sqrt{3}}{2} \frac{t_0 a} {\hbar}$ is the Fermi velocity of monolayer graphene and $a=2.46 \rm{\AA}$ is the lattice constant, to effectively account for the overestimation of the exchange by long-ranged Coulomb repulsion in a Hartree-Fock theory that misses out the screening effects of $\pi$ and $\sigma$ orbitals in graphene. 
%(Other types of screening approximations partially taking into account the screening effects such as a static Thomas-Fermi approximation for the exchange interaction do not change the qualitative picture presented in this paper.) 
(Simple screening models such as the static Thomas-Fermi approximation for the exchange interaction do not change the qualitative picture on the sublattice symmetry breaking presented in this paper.) 
The specific value $\alpha = 0.1$ is adopted to match the experimentally observed gap size in bilayer graphene \cite{freitag, Feldman2009, Veligura2012}.

To overcome computational challenges posed by the absence of analytic form of wavefunctions in multilayer graphene, we use the rotational transformation method \cite{Jang2015} in which the wavefunction at an arbitrary angle is obtained by a stacking dependent unitary transformation of the wavefunction at a specific angle. Moreover, we omit the inter-valley interaction which is negligibly small, and each one of the four spin/valley flavors are treated independently.
%%%%%%%%%%%%%%%%%%%%%%%%%%%%%%%%%%%%%%%%%%%%%%%%%%%%%%%%%%%%%%%%%%%%%%%%%%%%%%%%%%%%%%%%%%%%%%%%%%%%

%%%%%%%%%%%%%%%%%%%%%%%%%%%%%%%%%%%%%%%%%%%%%%%%%%%%%%%
\begin{figure*}[t]
\frame{\includegraphics[width=\linewidth]{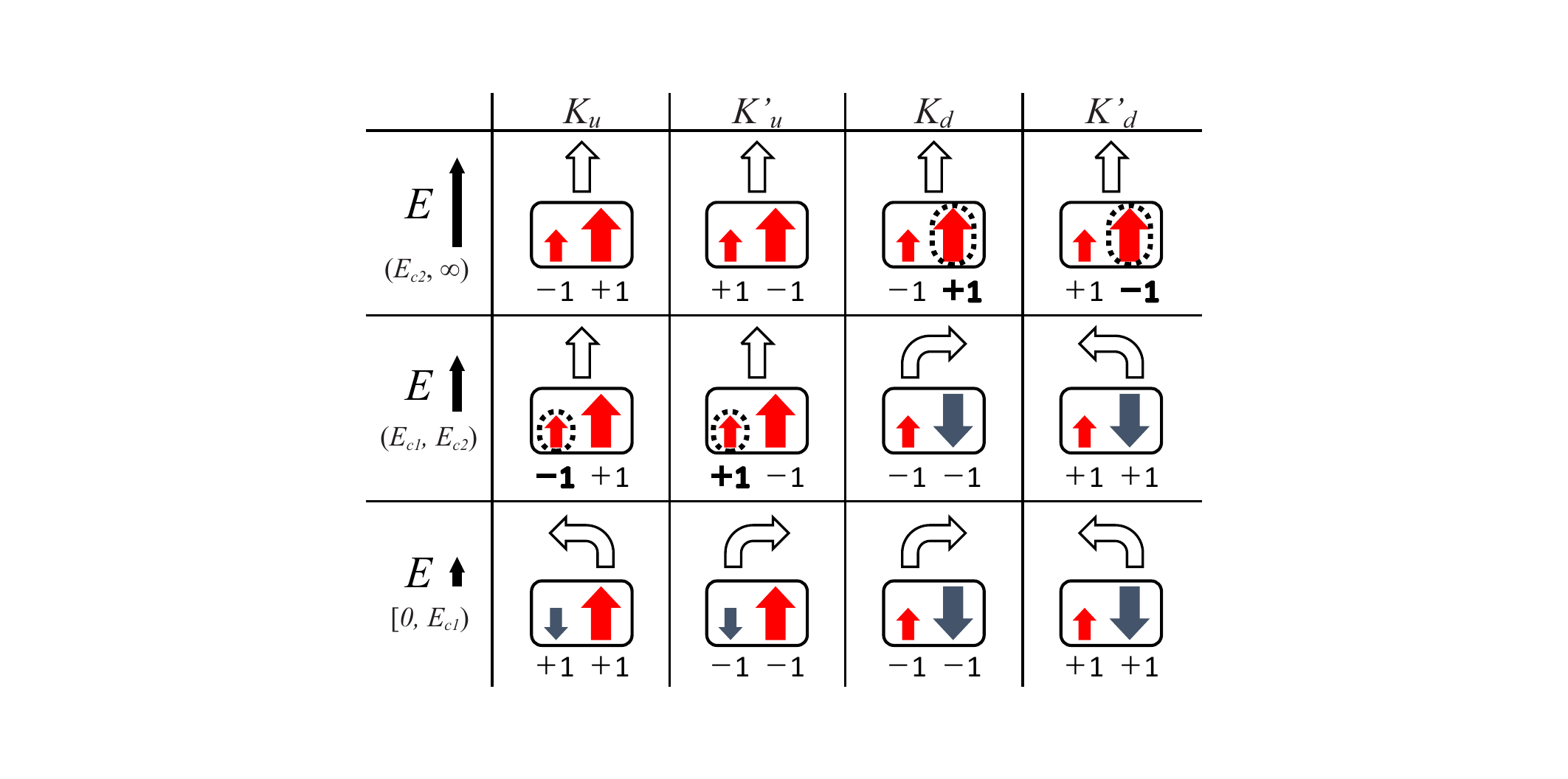}}
\caption{
%(a) Schematic picture of the ground-state configuration and corresponding spontaneous Hall effect at zero external electric field for three possible flavor antiferro states: LAF, QSH and QAH.
The evolution of the layer antiferromagnetic state under a perpendicular external electric field, keeping the flavor degeneracy of the system.
%Here, the antiferro character is imposed when external electric field is applied. 
Arrows in the square box and numbers below the box at each spin/valley flavor represent pseudospin polarizations and corresponding Chern numbers, respectively, whereas the arrows above the box indicate the corresponding net current directions expected in the Hall measurement. The change in the charge polarization by applying a perpendicular electric field is denoted by the dashed circle.
%Red colored arrows and numbers indicate the change in pseudospin polarization directions and corresponding Chern numbers when external electric field is applied.
} 
\label{fig:Hall_effect_picture}
\end{figure*}
%%%%%%%%%%%%%%%%%%%%%%%%%%%%%%%%%%%%%%%%%%%%%%%%%%%%%%%

\section{Results}
\subsection{Interaction-driven gapped phases in Bernal stacked tetralayer graphene}
The experimentally observed band gap in Bernal stacked tetralayer graphene suggests the presence of electron-electron interaction driven symmetry breaking \cite{Grushina2015,Nam2016}. 
Here, we show that the band gap opens due to interaction driven sublattice symmetry breaking and its internal structure consists of light-mass and heavy-mass band doublets whose charge densities polarize towards opposite out-of-plane directions.
A sufficiently strong perpendicular external electric field can flip their polarization directions in the order of increasing effective mass values. 

In Fig.~\ref{fig:ground_state}, we show a comparison of the non-interacting and Hartree-Fock energy band structures, and corresponding ground-state wavefunction amplitudes and charge polarizations near the Fermi energy. 
The electronic structure of Bernal stacked multilayer graphene can be understood from the chiral decomposition rules of arbitrarily stacked multilayers \cite{Min2008b,Min2008c} where the ABAB tetralayer is the simplest example involving more than one massive band. In the absence of electron-electron interactions, the low-energy band structure of ABAB stacking is described by two bilayer-like pseudospin doublets with different effective masses, whose wavefunctions near the Fermi energy are mainly localized at outer layer (1A,~4B) and inner layer (2B,~3A) sublattice sites that define the pseudospin basis for the light and heavy mass bands, respectively, as shown in Fig.~\ref{fig:ground_state}(c). 
In the presence of electron-electron interactions, the sublattice symmetry of the two-fold degenerate pseudospin doublets in the occupied bands is broken by transferring charge either from A to B sublattices or vice versa for both doublets (but not from A to B for one doublet and from B to A for other doublet), resulting in the gapped band structure with the same \emph{sublattice} polarization direction.

Sublattice symmetry breaking in tetralayer graphene can be considered as the generalization of the case of bilayer graphene system, which has one pseudospin (per spin and valley) whose direction is out-of-plane as a result of the electron-electron interactions \cite{Min2008a}. 
For each spin/valley flavor, the charge polarizations of the light-mass and heavy-mass band doublets can be represented as $(\downarrow,\Uparrow)$ or $(\uparrow,\Downarrow)$, where the first (second) arrow in the parenthesis denotes the charge polarization for the light (heavy) mass band. 
The charge polarizations for light and heavy bands point in opposite directions but 
towards the same sublattices, leading to same sign Chern numbers ($+1$,$+1$) or ($-1$,$-1$), as shown  in the Fig.~\ref{fig:ground_state}(d).
Note that the Chern number changes its sign at the opposite valley. We will discuss later on the states with the same polarization directions such as  $(\downarrow,\Downarrow)$ or $(\uparrow,\Uparrow)$ with a vanishing net Chern number for a single spin/valley flavor, which are possible in the presence of an external electric field.
Thus, we can expect a variety of ground states as a function of an external electric field, where different types of Hall conductivities can result depending on the polarization of light-mass and heavy-mass band doublets for each flavor. 

Similar to the discussions for pseudospin magnetism 
in bilayer graphene \cite{Min2008a} we can classify the different states into flavor antiferro, ferri, and ferro states.
Flavor antiferro states have two flavors in $(\downarrow,\Uparrow)$ configuration and the other two in $(\uparrow,\Downarrow)$ configuration at zero field, so that no net charge polarization exists. Flavor antiferro states can be further classified depending on their Hall conductivities \cite{Supplemental}. 
Flavor ferro states have all four flavors in the same pseudospin configuration. The flavor ferri states have one distinct flavor with respect to other three. Since the number of $(\downarrow,\Uparrow)$ and $(\uparrow,\Downarrow)$ configurations are different in flavor ferro and ferri states at zero field, non-zero net charge polarization exists for these states. 
From the Hartree energy cost considerations, the metastable states with the lowest total energy are expected to be flavor antiferro when there is no external electric field perpendicular to the graphene layers. 

%%%%%%%%%%%%%%%%%%%%%%%%%%%%%%%%%%%%%%%%%%%%%%%%%%%%%%%%%%%%%%%%%%%%%%%%%%%%%%%%%%%%%%%%%%%%%%%%%%%%

%%%%%%%%%%%%%%%%%%%%%%%%%%%%%%%%%%%%%%%%%%%%%%%%%%
\begin{figure*}[t]
\frame{\includegraphics[width=1	\linewidth]{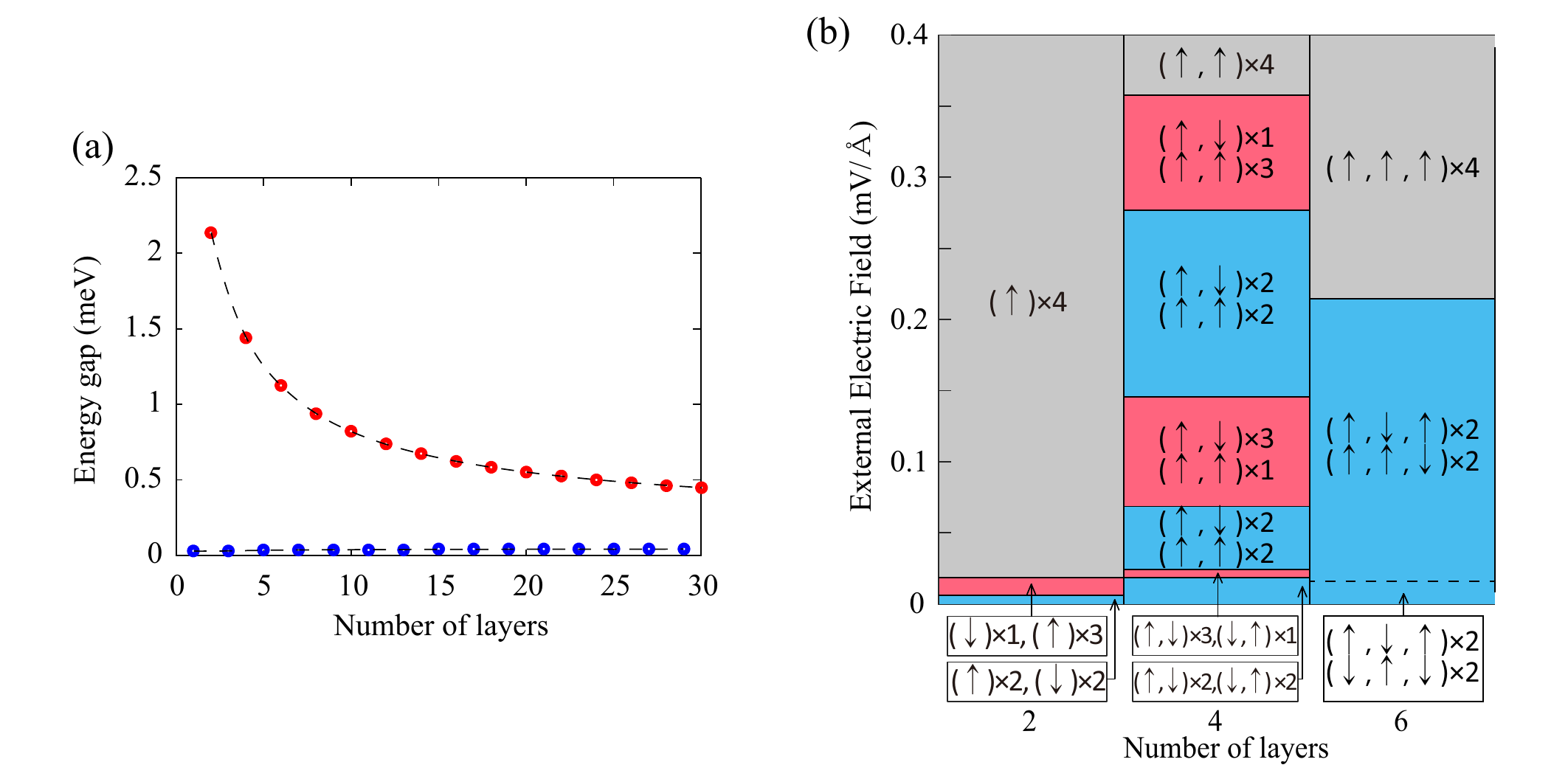}}
\caption{
(a) Energy band gap as a function of the number of layers in the absence of a perpendicular electric field. The energy gaps of even (odd) number of layers are denoted by red (blue) circles. (b) The lowest total energy states in the presence of electric field. Flavor antiferro, ferri, and ferro states are colored in blue, red, and gray, respectively. The pseudospin polarization directions are written in increasing effective mass order.} 
\label{fig:even_odd}
\end{figure*}
%%%%%%%%%%%%%%%%%%%%%%%%%%%%%%%%%%%%%%%%%%%%%%%%%%

\subsection{Electric field induced ``Hund's rule" and Hall effects}
%{\em Spontaneous quantum Hall states and ``Hund's rule"} ---
Now let us consider the effect of a perpendicular external electric field that can introduce a richer phase diagram. The presence of an electric field is able to reorganize the charge polarization of the sublattice pseudospins in each spin/valley flavor. 
%For clearer understanding for pseudospin configuration change under external electric field, we impose the system not to change its antiferro, ferri, ferro character under external electric field. The ground state under external field is discussed in the following section.
(Here we evolve a pseudospin configuration under an electric field without changing its antiferro, ferri or ferro character keeping the same flavor degeneracy. The lowest total energy state among them is discussed in the following section.)
We begin by considering the flavor antiferro state
consisting of $(\downarrow,\Uparrow)\times 2$ and $(\uparrow,\Downarrow)\times 2$ at zero field. 
When an external electric field is increased beyond the first critical field of $E_{\rm {c1}}= 0.025$ mV/$\rm{\AA}$, the polarization of the light-mass band changes its sign first due to smaller interaction-induced sublattice potential
compared with that for the heavy-mass band, resulting in the $(\uparrow,\Uparrow)\times 2$ and $(\uparrow,\Downarrow)\times 2$ configuration.
A second critical electric field of $E_{\rm {c2}}= 0.879$ mV/$\rm{\AA}$ is able to flip all pseudospins  leading to a ground state with four identical copies of the band doublets $(\uparrow,\Uparrow)\times 4$, and thus resulting in flavor ferro state.  
Figure ~\ref{fig:Hall_effect_picture} schematically illustrates this process and resulting transport properties in one of the flavor antiferro states, the layer antiferromagnetic phase, assuming a spin dependent but valley independent sublattice potential \cite{Supplemental}.
%within the flavor antiferro configuration. 
Thus, polarizations of the pseudospin doublets arising from the interaction induced sublattice symmetry breaking 
are aligned by the external electric field in the order of increasing effective mass. 
It can be shown that this simple ``Hund's rule" type pseudospin ordering applies also to flavor ferri and ferro states, whose detailed discussion can be found in the Supplemental Material \cite{Supplemental}.

%%%%%%%%%%%%%%%%%%%%%%%%%%%%%%%%%%%%%%%%%%%%%%%%%%%%%%%%%%%%%%%%%%%%%%%%%%%%%%%%%%%%%%%%%%%%%%%%%%%%
\subsection{Generalization to thicker multilayer stacks}
In Bernal stacked multilayer graphene beyond tetralayer, there are several additional factors that influence electronic structure near the Fermi level
due to the increased number of pseudospin doublets and their interactions.
Here, we intend to provide a qualitative picture of the electronic structure expected in Bernal stacked multilayer graphene in the presence of electron-electron interactions and perpendicular external electric fields 
using a minimal continuum model for the band Hamiltonian. 
%%%%%%%%%%%%%%%%%%%%%%%%%%%%%%%%%%%%%%%%%%%%%%%%%%%
%\begin{figure}[t]
%\includegraphics[width=1\linewidth]{fig4.eps}
%\caption{
%The lowest total energy states in the presence of electric field. Antiferro, ferri, and ferro states are colored in blue, red, and gray, respectively. The pseudospin polarization directions are written in increasing effective mass order.
%} 
%\label{fig:lowest_total}
%\end{figure}
%%%%%%%%%%%%%%%%%%%%%%%%%%%%%%%%%%%%%%%%%%%%%%%%%%%

In Bernal stacked multilayer graphene, the low-energy effective theory is described by a set of bilayer-like doublets for even-number of layers, while an additional monolayer-like doublet is found for odd-number of layers. The major difference in the energy gap between even- and odd-layer graphenes originates from the existence of the monolayer-like doublet in odd layered multilayers~\cite{Guinea2006, Latil2006, Partoens2007, Min2008b}. Unlike bilayer-like doublets, monolayer-like doublets are much more robust to the interaction-induced sublattice symmetry breaking and tend to remain gapless \cite{Min2008a}, thus the gaps of odd-layer graphenes are much smaller than those of even-layer graphenes. 

%In Fig.~\ref{fig:even_odd}(a) we show the energy band gap as a function of the number of layers in the absence of electric field. For odd number of layers (but not for a single layer), the energy gap remains finite and saturates as the number of layers increases. For even number of layers, the energy gap opens due to interactions but it also saturates as the number of layers increases. The finite gap for odd layers and the saturation gap for even layers arise from the difference between the intralayer and interlayer exchange interactions, which vanish when we set the zero interlayer separation. It is important to note that the remnant gaps are due to simplification in the minimal model and expected to be closed when remote hopping terms and screening are considered. Details of the limitation of the minimal model will be discussed in the discussion section.

In Fig.~\ref{fig:even_odd} (a), we show the energy band gap as a function of the number of layers in the absence of electric field. For even number of layers the energy gap opens due to interactions and decreases as the number of layers increases, whereas for odd number of layers the energy gap almost remains closed. Note that the energy gap for odd number of layers is not exactly zero (except for a single layer) and the energy gap for both odd and even layers saturate as the number of layers increases.
%, which arise due to the difference between the intralayer and interlayer exchange interactions. 
It is important to note that these remnant gaps are due to simplification in the minimal model and expected to be closed when remote hopping terms and screening are considered. 

%In Fig.~\ref{fig:even_odd}(a) we show the energy band gap as a function of the number of layers in the absence of electric field. For odd number of layers (but not for a single layer), the energy gap remains finite and saturates as the number of layers increases. For even number of layers, the energy gap opens due to interactions but it also saturates as the number of layers increases. The finite gap for odd layers and the saturation gap for even layers arise from the difference between the intralayer and interlayer exchange interactions, which vanish when we set the zero interlayer separation. It is important to note that the remnant gaps are due to simplification in the minimal model and expected to be closed when remote hopping terms and screening are considered. Details of the limitation of the minimal model will be discussed in the discussion section.

%%%%%%%%%%%%%%%%%%%%%%%%%%%%%%%%%%%%%%%%%%%%%%%% 
 
Restricting our attention to even-layer graphene, we summarize in Fig.~\ref{fig:even_odd} (b) the effect of a perpendicular electric field in the ground-state configurations.
%In the case of bilayer and tetralayer graphene, the lowest total energy state varies from antiferro via ferri to ferro irrespective of the interaction strength $\alpha$. For six or more layer graphene, however, due to the small difference in total energy between metastable states, the detailed ground-state configurations depend on the value of $\alpha$. 
In general, the lowest total energy state varies from a flavor antiferro state with zero net charge polarization via a partially polarized state, and eventually to fully polarized flavor ferro state.
Interestingly, for an appropriate external electric field range, the flavor ferri state (or ``All" state \cite{Zhang2011, Zhang2015, jung2011}), 
which exhibits non-zero Hall conductivities of all flavors
can be achieved not only in rhombohedral but also in Bernal stacked multilayer graphene. 
Since the total energies are almost degenerate, we expect that domains of different pseudospin configurations will form in a disordered sample \cite{Li2014}. 
Considering the large number of pseudospin flavors in tetralayers and beyond, it is also expected that a large variety of topological domain walls
will arise at the interface between the ordered pseudospin domains.
%Because of multiple pseudospins for each spin/valley flavor in tetralayer and beyond, rich topological domain wall modes can appear between domains with different topological character arising from various interaction-induced pseudospin configurations.

%%%%%%%%%%%%%%%%%%%%%%%%%%%%%%%%%%%%%%%%%%%%%%%%%%%%%%%%%%%%%%%%%%%%%%%%%%%%%%%%%%%%%%%%%%%%%%%%%%%%
\section{Discussion}
We identified the structure of the electron-electron interaction driven ordered phases in Bernal stacked multilayer graphene based on the polarization of the pseudospin doublets belonging to electronic bands with distinct effective masses.
Our analysis rests on a number of simplifying assumptions such as: neglect of remote hopping terms and the energy difference between the dimer and non-dimer sites $\Delta$, and the absence of screening and correlations in our interaction model. 
In our minimal model for the band Hamiltonian, only the nearest intralayer and interlayer hopping is considered for simplicity in order to conserve the rotational symmetry of the Hamiltonian. As the number of layers becomes larger, however, the remote hopping terms cannot be omitted 
for an accurate description of the band structure. Each remote hopping term plays a different role in multilayer graphene, but in general, it distorts the chiral character of the low energy band near the $K$ or $K'$ point reducing the density of states near the Fermi energy. 
Since the energy gap originates from the interplay of chirality and electron-electron interaction, the energy gap is expected to become smaller when the remote hopping terms are considered. Once the energy gap is closed or becomes narrower, the screening effect due to the Coulomb interaction begins to play a significant role, in a particularly notable manner for odd-layer graphenes. 
%In even-layer graphenes, however, when the energy gap is the dominant energy scale and the effect of remote hopping terms are small, the basic picture presented in this paper remains valid at least qualitatively. 
For even-layer graphenes, larger interaction induced gaps open.
%it is possible that the interaction induced energy gap opens. 
When the gap size sets the dominant energy scale relative to the remote hopping energies, the basic picture presented in this paper should be valid at least qualitatively. 
It has been proposed % suggested
that the electron-electron interactions will induce strains that suppress the remote interlayer coupling terms such as the $\gamma_2$ hopping \cite{Nam2016}.  
%justifying the use of the minimal model. 
The assumption of weakened remote hopping terms in few layer systems would make the minimal model an adequate ground for the analysis of interaction effects.
%
%The assumption that the remote hopping terms are weakened weakened remote hopping terms should make adequate an analysis based
%on the minimal model picture.
%the minimal model analysis picture drawn from the minimal few layer models should provide 
%the analysis of interactions in the minimal model for few-layers 
%and we expect that the inclusion of the remote hopping terms do not change the basic picture drawn from the minimal model for few-layer samples.
%As the number of layers increases, however, the energy gap should decreases, and the remote hopping terms and screening should lead to a closure of the band gap as the system eventually becomes graphite.
As the number of layers increases towards the graphite limit, however, it is expected that the energy gap should show a progressive decrease until it eventually closes.
%due to an enhanced role of remote hopping terms and screening of electron-electron interaction.

%In even-layer graphenes, the energy gap is the dominant energy scale, thus when the gap is large enough, the effect of other energy scales associated with remote hopping terms could be neglibible and the basic picture presented in this paper remains valid at least qualitatively. Furthermore, it has been argued that the remote hopping terms in multilayer graphene could be different from those in graphite and suppressed by interactions \cite{Nam2016} justifying the use of the minimal model. Thus we expect that the remote hopping terms do not change the basic picture drawn from the minimal model.

%Once the distortion is not negligible, the minimal model is no more valid. Among the remote hopping terms, the one that makes the band structure most differently from that of minimal model is $\gamma_{2}$, the next-nearest interlayer hopping between non-dimer sites. The commonly accepted value of $\gamma_{2}$ obtained from graphite studies is $-20$ meV, which is much larger than the energy gaps of Bernal-stacked multilayer graphenes. However, a recent study \cite{Nam2016} argued that there is possibility of the suppression of $\gamma_{2}$ in few-layer graphenes resulting from the electron-electron interaction and doping.

In summary, we provide a simple and comprehensive picture for the interaction induced ordered states in Bernal stacked multilayer graphene. We analyze the ground-state configurations and associated Hall conductivities 
that can result from the combined presence of electron-electron interactions and perpendicular external fields
that could serve as guidance to future experiments.
%in the presence of electron-electron interactions and perpendicular external electric fields, which could be useful guidance for future experimental studies on multilayer graphene in the presence of interaction-induced ordered phases.

%%%%%%%%%%%%%%%%%%%%%%%%%%%%%%%%%%%%%%%%%%%%%%%%%%%%%%%%%%%%%%%%%%%%%%%%%%%%%%%%%%%%%%%%%%%%%%%%%%%%

\section{Acknowledgments}
Authors thank A. H. MacDonald, F. Zhang and J.-J. Su for helpful discussions.
This research was supported by Basic Science Research Program through the National Research Foundation of Korea (NRF) funded by the Ministry of Education under Grant No. 2015R1D1A1A01058071. J.J. acknowledges financial support from the Korean NRF through the grant NRF-2016R1A2B4010105. C.Y. acknowledges the financial support of the NRF grant funded by the Korean Government (NRF-2016H1A2A1907780).

%%%%%%%%%%%%%%%%%%%%%%%%%%%%%%%%%%%%%%%%%%%%%%%%%%

%%%%%%%%%%%%%%%%%%%%%%%%%%%%%%%%%%%%%%%%%%%%%%%%%%
\newpage
\pagenumbering{gobble}

\begin{figure}[htp]
\includegraphics[page=1,trim = 19mm 17mm 17mm 17mm]{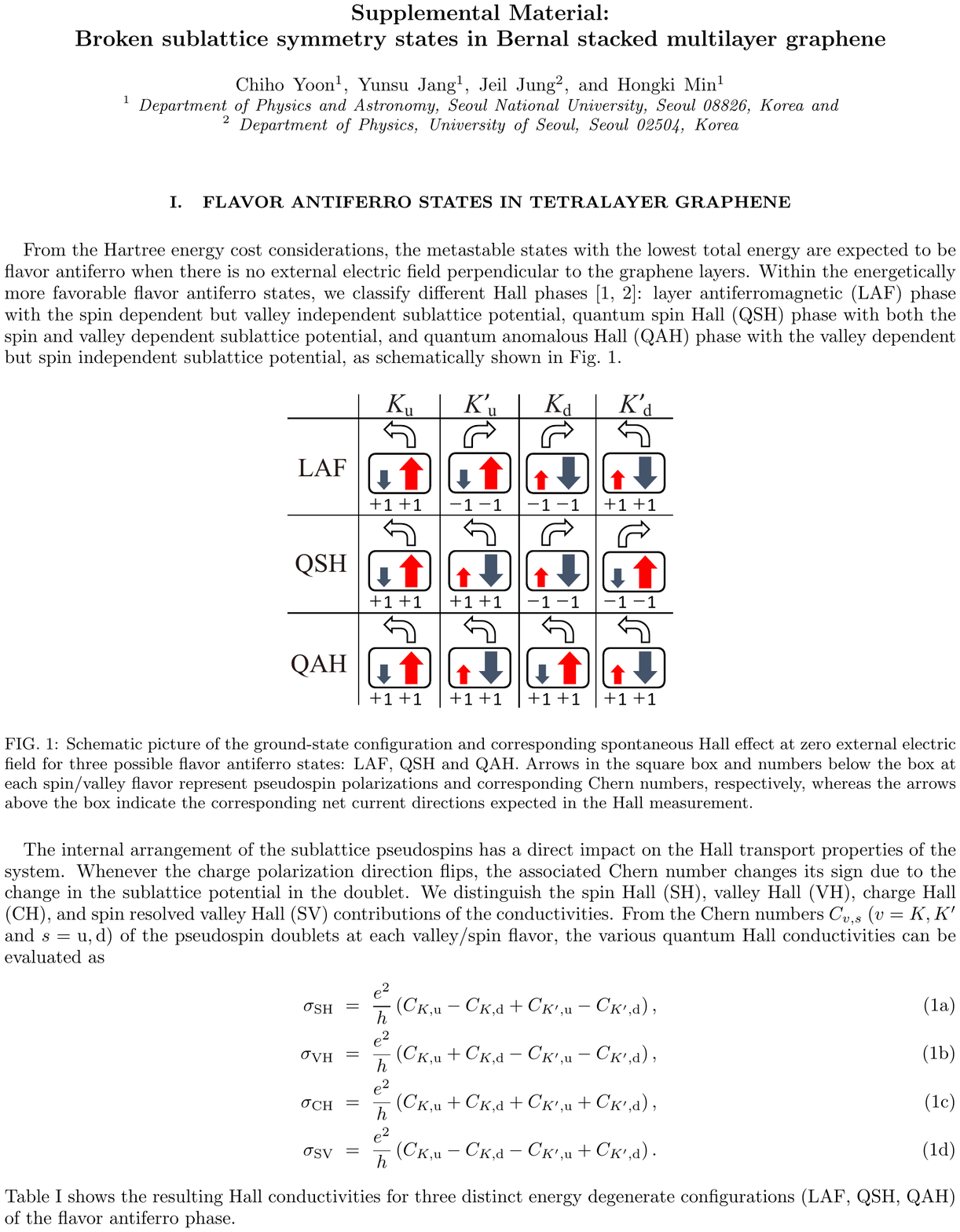}

\end{figure}

\newpage

\begin{figure}[htp]
  \includegraphics[page=2,trim = 19mm 17mm 17mm 17mm]{supplemental.pdf}

\end{figure}

\newpage

\begin{figure}[htp]
  \includegraphics[page=3,trim = 19mm 17mm 17mm 17mm ]{supplemental.pdf}

\end{figure}

\newpage

\begin{figure}[htp]
  \includegraphics[page=4,trim = 19mm 17mm 17mm 17mm ]{supplemental.pdf}

\end{figure}

\newpage

\begin{figure}[htp]
  \includegraphics[page=5,trim = 19mm 17mm 17mm 17mm ]{supplemental.pdf}

\end{figure}

\end{document}